\documentclass{article}
\usepackage{spconf, amsmath, amsfonts, graphicx}
\usepackage{booktabs} 
\usepackage{bm,  algorithm, algorithmic}

\usepackage{dblfloatfix}  

\DeclareMathOperator*{\argmin}{arg\,min}

\newtheorem{theorem}{Theorem}

\newtheorem{lemma}{Lemma}

\title{A Dimension-Independent discriminant between distributions}
\name{Salimeh Yasaei Sekeh, Brandon Oselio, Alfred O. Hero III\thanks{This work was partially supported by ARO grant W911NF-15-1-0479.}}
\address{Department of Electrical Engineering and Computer Science\\
University of Michigan\\
1301 Beal Ave, Ann Arbor, MI, 48109, USA}
\begin{document}
\ninept
\maketitle
\begin{abstract}

Henze-Penrose divergence is a non-parametric divergence measure that can be used to estimate a bound on the Bayes error in a binary classification problem. In this paper, we show that a cross-match statistic based on optimal weighted matching can be used to directly estimate Henze-Penrose divergence. Unlike an earlier approach based on the Friedman-Rafsky minimal spanning tree statistic, the proposed method is dimension-independent. The new approach is evaluated using simulation and applied to real datasets to obtain Bayes error estimates.
\end{abstract}

\def\BX{\mathbf{X}}
\def\BY{\mathbf{Y}}
\def\bx{\mathbf{x}}
\def\R{\mathcal{R}}
\def\diy{\displaystyle}
\def\rd{\mathrm{d}}
\section{Introduction}
\label{sec:intro}

Many information theoretic measures have been applied to measure the discrimination between probability density functions. They have been used in various applications in signal processing, classification, image registration, clustering and structure learning, see \cite{Guorongetal1996,ViolaWilliam1997,HamzaKrim2003,MNYSH2017}. A special class of divergence measures, called $f$-divergences have the property that the divergence functional $f$ is convex and $f(1)=0$. Among the different divergence functions belonging to the $f$-divergence family, \cite{AliS1996,Csisz1967} the Henze-Penrose (HP) divergence has been of great interest due to its application to binary classification, in particular to bound the Bayes error rate.


Let $\bx_1,\bx_2,...,\bx_N \in \R^d$ be realizations of random vector $\BX$ and class labels $y\in\{0,1\}$, with prior probabilities $c_0=P(y=0)$ and $c_1=P(y=1)$, such that $c_0+c_1=1$. Given conditional distributions $p_0(\bx)$ and $p_1(\bx)$, the Bayes error rate  is given by

\begin{equation}\label{BER}
\epsilon=\diy\int_{\R^d}\min\big\{c_0p_0(\bx),c_1p_1(\bx)\big\}\rd \bx.
\end{equation}
The Bayes error rate is the expected risk for the Bayes classifier, which assigns a given feature vector $\bx$ to the class with the highest posterior probability, and is the lowest possible error rate of any classifier for a particular joint distribution. It is thus a reasonable measure for assessing the intrinsic difficulty of a particular classification problem. By estimating and bounding this value, we can then have a better understanding of the problem difficulty, which allows the user to make more informed decisions.

We define the HP-divergence between $p_0$ and $p_1$, $D_{c}(p_0,p_1)$ by
\begin{equation}\label{Def:HP}
\diy\frac{1}{4c_0c_1}\left[\int_{\R^d}\diy\frac{\big(c_0p_0(\bx)-c_1p_1(\bx)\big)^2}{c_0p_0(\bx)+c_1p_1(\bx)}\;\rd\bx-(c_0-c_1)^2\right].
\end{equation}
Note that for all $c_0$ and $c_1$, $0\leq D_c(p_0,p_1)\leq 1$ and when $p_0=p_1$ the HP-divergence becomes zero. 


The authors of \cite{Berishaetal2016} showed that HP-divergence yields tighter bounds on the Bayes error rate $\epsilon$, given in (\ref{BER}), than those based on the Bhattacharya distance, \cite{Bha}. In particular, the following bound on the Bayes error rate holds:
\begin{equation}
\frac{1}{2} - \frac{1}{2}\sqrt{u_c(p_0, p_1)} \le \epsilon \le \frac{1}{2} - \frac{1}{2}u_c(p_0,p_1),
\end{equation}
where $u_c(p_0,p_1)=4c_0c_1D_c(p_0,p_1)+(c_0-c_1)^2$.

In this paper we propose a new direct estimator for HP-divergence using a statistic based on optimal weighted matching \cite{Rosenbaum2005}. Matching for general graphs is a combinatorial optimization problem that can be solved in polynomial time. In \cite{Rosenbaum2005}, the optimal weighted matching was used to find a statistical test for equal posterior distributions using the cross match statistic. We demonstrate that the same statistic described in that series of papers can be utilized to estimate HP-divergence. We emphasize that the proposed weighted matching estimator is completely different from weighted $K$-NN graph estimators. 

The rest of the paper is organized as follows. Section 2 briefly describes related work on HP-divergence and optimal weighted matching. Section~3 defines the cross-match statistic, and in Section~4 we prove that the cross-match statistic approximately tends to the HP-divergence when samples sizes of two classes increases simultaneously in a specific regime. Section~5 shows sets of simulations for our proposed method and compares the Friedman-Rafsky (FR) and cross-match estimators experimentally, and we estimate the Bayes error rate on a few real datasets. Finally, Section~6 concludes the paper.

\begin{figure*}[!t]
\centering{}
  \includegraphics[width=\linewidth]{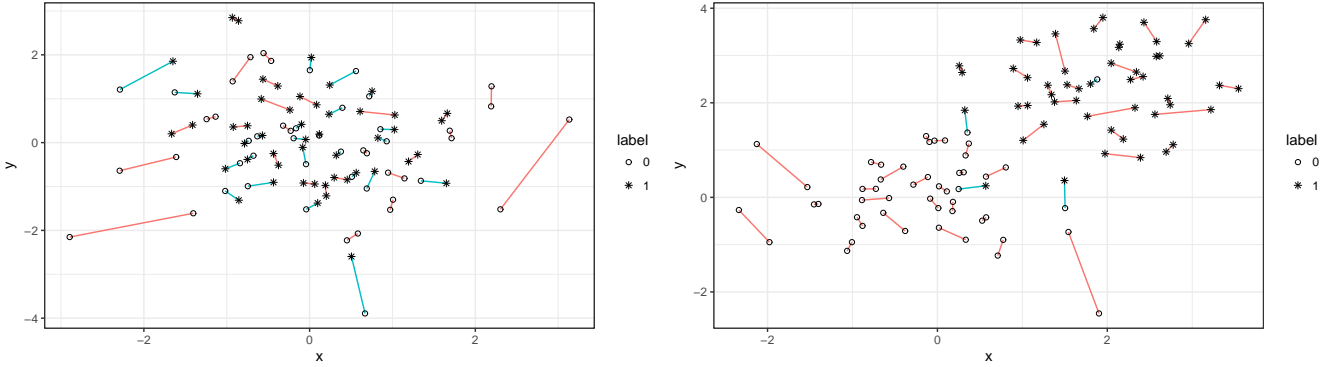}
  \caption{{\small An example of the cross-match statistics for two cases $p_0=p_1$ (left-generated from standard Gaussian distributions) and $p_0\ne p_1$ (right-Generated from Gaussian distributions with means $[0,0]$, $[2,2]$). The total number of blue edges is the cross match statistics.}}
  \label{fig:matching_ex}
\end{figure*}
\section{RElated Work}
\label{sec:related_work}
Several estimators for HP-divergence have been proposed in the literature: Plug-in estimates were introduced in \cite{Scricharanetal2012} and later have been studied more in \cite{MoonHero2014,MoonHero2016,Moonetal2016}. Plug-in approaches estimate the underlying distribution function and then plug this value into the divergence function. The drawback with the plug-in estimates is that these methods are not accurate near support boundaries and are also more computationally complex. There have been a number of attempts to non-parametrically approximate divergence measures using graph-based algorithms such as minimal spanning tree (MST), \cite{Yukish,AldousSteel1992} and $k$-nearest neighbors graphs ($k$-NNG), \cite{Beardwoodetal1959}.

One of the most common direct estimators is based on Friedman-Rafsky (FR) multivariate test statistic \cite{FR}. This approach is constructed from the MST on the concatenated data set drawn from sufficiently smooth probability densities. Henze and Penrose \cite{HP} showed that the FR test is consistent against all alternatives. Therefore,  the HP-divergence has the appealing property that there exists an asymptotically consistent direct estimator in terms of the FR test statistic, see \cite{HP, BerishaHero2015,Berishaetal2016}. The variance of the FR test statistic under the assumption of equal distributions depends on the dimension of the data $d$, which may be unknown, especially when the support of the densities is a common but unknown lower dimensional manifold.

Optimal weighted matching is a well studied combinatorial optimization problem  \cite{PapadimitriouSteiglitz1982}. It has been used extensively in operations engineering. Previous statistical work using weighted matching have derived useful applications of the cross-match test statistic in fields like biological networks \cite{Rosenbaum2005,LuRosenbaum2004}.

\section{The Cross-Match test statistic}
\label{sec:algorithm}
\def\br{\mathbf{r}}
\def\BR{\mathbf{R}}
\def\BS{\mathbf{S}}
\def\by{\mathbf{y}}
Consider $N$ i.i.d. samples $\mathcal{X}_{N}=\{\bx_1,\ldots,\bx_N\},~\bx_i\in\R^{d}$ and corresponding labels $y_i\in\{0,1\}$. Define $\by=(y_1,\dots,y_N)$, and further $m=\sum\limits_{l=1}^{N} y_l$, and $n=N-m$, so that $m$ is the number of samples in $\bx$ with class $1$, and  $n$ is the number with class label $0$. Further, we create $D$, a $N\times N$ Euclidean distance matrix, with $D_{ij} = ||\bx_i - \bx_j||$. Without loss of generality, we assume $N$ is even, as we can always add a `ghost point' $\bx_{N+1}$, where $D_{i N+1} = 0, \forall i$. In the following, we consider a complete weighted graph $G = (V,E,D)$, with the vertices $V = 1,\ldots,N$ representing the sample points $\bx_1,\ldots,\bx_N$, edges $E = \{\{i, j\},~i, j \in V\}$, and weights for each edge $\{i,j\}$ as $D_{ij}$.

A complete matching $M \subset E$ on a weighted graph is a set of edges such that no two edges in $M$ share a common vertex, and every vertex is used in the matching. The complete minimum weighted matching $M^*$ is defined as the matching on $G$ such that $M^* = \argmin_{M} \sum_{{i, j} \in M} D_{ij}$.  We note that this is similar to the FR test \cite{FR}, which uses the same matrix $D$ to find the minimal spanning tree. The FR test statistic is the total number of edges in the $D$-based MST connecting different labeled nodes.  

Using this matching, we find the {\textit{cross-match statistic}}, $\mathcal{A}(\mathcal{X}_N)$ which is the number of edges that match dichotomous samples, i.e. samples with different class labels, that is 
\begin{equation}
\mathcal{A}(\mathcal{X}_N)=\diy\sum_{\{i, j\} \in M^*}\Big(y_{i}(1-y_{j})+(1-y_{i})y_{j}\Big).
\end{equation}


In Figure \ref{fig:matching_ex} we show two numerical examples. The left plot shows samples from two equal distributions, and right plot shows samples from differing distributions. Qualitatively, we see that $\mathcal{A}$ is much greater for the equivalent distributions than for the differing distributions, because the optimal matching tries to reduce long distances, which will reduce the number of edges between differing distributions.

In Proposition 1 in \cite{Rosenbaum2005}, under the assumption of equal distributions, the expectation and variance of $\mathcal{A}(\bx)$ are derived:
\begin{equation}
E[\mathcal{A}]=\diy\frac{mn}{N-1},\;\; Var[\mathcal{A}]=\diy\frac{2n(n-1)m(m-1)}{(N-3)(N-1)^2}.
\end{equation}
We note that the mean and variance of the cross-match statistic under equal distributions are dimension-independent, but this is not true for the FR statistic, whose variance is dependent on the degrees of the MST. The maximal degrees of the MST is in fact dependent on the dimension $d$ of the underlying samples, e.g., the MST has maximal degree 4 in $d=2$ dimensions while its maximal degree is known to be between 13 or 14 in $3$ dimensions \cite{robins1994maximum}. This dependence causes the FR statistic to perform poorly in higher dimensions. In Section \ref{sec:experiments} we perform a set of experiments where dimension varies to demonstrate the advantage of the cross-match statistic over the FR statistic.

\def\BZ{\mathbf{Z}}
\def\bbE{\mathbb{E}}
\def\diy{\displaystyle}
\def\bx{\mathbf{x}}
\def\rd{\mathrm{d}}
\def\BU{\mathbf{U}}
\def\bu{\mathbf{u}}

\section{Hp-divergence Estimation}
\label{sec:mainresult}
Here we introduce the cross-match statistic as an estimate of the HP-divergence given in (\ref{Def:HP}). Assume that we have two sets of samples $\mathcal{X}_m=\{\BX_1,\dots,\BX_m\}$ and $\mathcal{U}_n=\{\BU_1,\dots,\BU_n\}$ with two different labels. In order to show asymptotic convergence to HP-divergence, we make the following assumption regarding the cross-match statistic (similar to Lemma 1 in~\cite{HP}).\\
{\bf Assumption 1:} For disjoint sets $\mathcal{X}_m$, $\mathcal{U}_n$ and $\{s,t\}$ we have
\begin{equation}\label{lem1:eq1}
\Big|\mathcal{A}(\mathcal{X}_m\cup\{s,t\}\cup\mathcal{U}_n)-\mathcal{A}(\mathcal{X}_m\cup\mathcal{U}_n)\Big|\leq k_d. 
\end{equation}
where $k_d$ is a constant that may depend on $d$.
This means that even if the optimal matching changes a great deal, the number of edges that are between the two samples is still approximately the same. 

\begin{figure}[h]
  \includegraphics[width=3.5in]{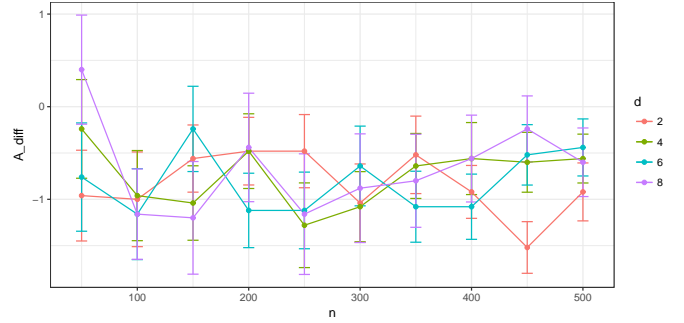}
  \caption{The cross-match statistics difference with error bars at the standard deviation from 50 trials for the Gaussian samples by adding two points.}
  \label{fig2}
\end{figure}

We empirically check this assumption in Figure~\ref{fig2}. We generate two sets of $d$-dimensional samples from standard Gaussian with mean $\mu_0=[0]_d$, $\mu_1=[1]_d$ and $\Sigma_0 = \Sigma_1 = I_d$ for $d=2,4,6,8$. We plot the difference in cross-match statistic when adding two points (labeled by $\mathcal{A}_{\mathrm{diff}}$), and perform this test over varying sample size. We see that $\mathcal{A}$ does not vary significantly when adding a new sample in the tested cases. 

\begin{lemma}\label{lem2}
Let $g:\mathcal{R}^d\times\mathcal{R}^d\rightarrow [0,1]$ be a symmetric and measurable function, such that for almost every $\bx\in\mathcal{R}^d$, $g(\bx,.)$ is measurable with $\bx$ a Lebesgue point of the functions $p(.)g(\bx,.)$ and $p(.)$. For each $N$, let $\BZ_1^N, \BZ_2^N,\dots, \BZ_N^N$ be independent $d$-dimensional variables with common density function $p_N$ convergent to $p$ as $N\rightarrow \infty$ and set $\mathcal{Z}_N=\{\BZ_1^N,\dots,\BZ_N^N\}$. Consider the complete minimum weighted matching $M^*$ on $\mathcal{Z}_N$. Then 
\begin{equation}\label{lem2:eq1}\begin{array}{l}
\diy\lim\limits_{N\rightarrow \infty}N^{-1}E\mathop{\sum\sum}_{1\leq i<j\leq N} g(\BZ_i^N,\BZ_j^N)\mathbf{1}\big\{(\BZ_i^N,\BZ_j^N)\in M^*(\mathcal{Z}_N)\big\}\\
\\
\qquad=\diy\frac{1}{2}\int_{\mathcal{R}^d} g(\bx,\bx) \; p(\bx). 
\end{array}\end{equation}
\end{lemma}
{\bf Proof:} For given $\bx$ in a subset $\mathcal{S}\in\mathcal{R}^d$, the degree of vertex $\bx$ in $M^*(\mathcal{S})$ is one.  Let $\bx$ be a Lebesgue point of $p(.)$ and $p(.)g(\bx,.)$ and $\mathcal{Z}_N^{\bx}$ be the point process $\{\bx,\BZ_2^N,\BZ_3^N,\dots,\BZ_N^N\}$. Let $\mathcal{B}(\bx,r)=\big\{\by:\|\by-\bx\|\leq r\big\}$. Therefore, we can write
\begin{equation}\label{lem2:eq1}\begin{array}{l}
E\diy\sum\limits_{j=2}^N\big|g(\bx, \BZ_j^N)-g(\bx,\bx)\big|\mathbf{1}\big\{\BZ_j^N\in \mathcal{B}(\bx,N^{-1/d})\big\}\\
\quad\qquad =(N-1)\diy\int_{\mathcal{B}(\bx,N^{-1/d})}\big|g(\bx,\by)-g(\bx,\bx)\big|p_N(\by)\;\rd\by\\
=(N-1)\diy\int_{\mathcal{B}(\bx,N^{-1/d})}\big|g(\bx,\by)p_N(\by)-h(\bx,\bx)p_N(\bx)\\
\\
\qquad \qquad +\diy g(\bx,\bx)(p_N(\bx)-p_N(\by)\big|\;\rd\by,
\end{array}\end{equation}
Since $\bx$ is a Lebesgue point of $p_N$ and $g(\bx,.)P_{N}(.)$ then (\ref{lem2:eq1}) tends to zero. Note that the degree of vertex in $M^*(\mathcal{Z}_N^{\bx})$ is one. For almost all $\bx$, 
\begin{equation}\label{lem2:eq3}
E\diy\sum_{j=2}^N g(\bx,\BZ_j^N)\mathbf{1}\big\{(\bx,\BZ_j^N)\in M^*(\mathcal{Z}_N^x)\big\}=g(\bx,\bx)+o(1).
\end{equation}
The function $g$ has range $[0,1]$ so the left hand side of (\ref{lem2:eq3}) is bounded by one. By the dominated convergence theorem 
\begin{equation}\label{lem2:eq2}\begin{array}{l}
N^{-1} E\mathop{\sum\sum}_{1\leq i<j\leq N} g(\BZ_i^N,\BZ_j^N)\mathbf{1}\big\{(\BZ_i^N,\BZ_j^N)\in M^*(\mathbf{Z}_N)\big\}\\
\\
\quad=\diy\frac{1}{2}E\sum\limits_{j=2}^N g(\BZ_1^N,\BZ_j^N)\mathbf{1}\big\{(\BZ_1^N,\BZ_j^N)\in M^*(\mathbf{Z}_N)\big\}\\
\quad=\diy\frac{1}{2}\int_{\bx} p_N(\bx)E\sum\limits_{j=2}^N g(\bx,\BZ_i^N)\mathbf{1}\big\{(\bx,\BZ_j^N)\in M^*(\mathbf{Z}_N)\big\}.
\end{array}\end{equation}
The last line in (\ref{lem2:eq2}) tends to right hand side of (\ref{lem2:eq1}). \hfill$\square$

The following theorem proves the direct estimate of HP-divergence based on $\mathcal{A}(\mathcal{X}_N)$. Due to space limitations only an outline of the proof is given.
\begin{theorem}\label{thm1} As $m\rightarrow \infty$ and $n\rightarrow \infty$ such that $m/N\rightarrow c_1$ and $\diy n/N\rightarrow c_0$, where $N=m+n$. Denote $\mathcal{A}_{m,n}:=\mathcal{A}(\mathcal{X}_m \cup \mathcal{U}_n)$ the cross-match statistic given by the optimal weighted matching over $\mathcal{X}_m$ and $\mathcal{U}_n$. Then under Assumption 1 we have
\begin{equation}\label{main:result}
1-\left(\frac{N}{m\;n}\right)\mathcal{A}_{m,n}\rightarrow D_c(p_0,p_1), \;\; \hbox{a.s.}
\end{equation}
\end{theorem}
{\bf Proof:} The proof shares some similarity with the FR convergence proof of the HP-divergence in \cite{HP}. The primary difference lies in handling the difference between the cross-match statistic when nodes are added, i.e. (\ref{lem1:eq1}). We use Lemma \ref{lem2} and Poissonization to prove (\ref{main:result}). 
Let $M_m$ and $N_n$ be Poisson variables with mean $m$ and $n$ such that $m+n$ is even and independent of one another and of $\BX
_i$ and $\BU_j$. Let $\mathcal{X}'_m$ and $\mathcal{U}'_n$ be the Poisson processes $\big\{\BX_1,\dots,\BX_{M_m}\big\}$ and $\big\{\BU_1,\dots,\BU_{N_n}\big\}$, respectively. Set $\mathcal{A}'_{m.n}$ = $\mathcal{A}(\mathcal{X}'_m\cup\mathcal{U}'_n)$, the cross-match statistic. By  (\ref{lem1:eq1}), we have 
\begin{equation}
\Big|\mathcal{A}'_{m,n}-\mathcal{A}_{m,n}\Big|\leq k_d\ \big(|M_m-m|+|N_n-n|\big). 
\end{equation}
Note that $(m+n)^{-1}E\big|\mathcal{A}'_{m,n}-\mathcal{A}_{m,n}|\rightarrow 0$. Poissonization makes the identities of the points of $\mathcal{X}'_m\cup\mathcal{U}'_n$ conditionally independent, given their positions. For each $m$ and $n$ let $\BZ_1^{m,n},\BZ_2^{m,n},\dots $ be independent discrete variables with common density $p_{m,n}(\bx)=(m p_0(\bx)+n p_1(\bx))/(m+n)$. Let $W_{m,n}$ be an independent Poisson variable with even valued mean $(m+n)$. Let $\mathcal{Z}'_{m,n}=\{\BZ_1^{m,n},\dots,\BZ_{W_{m,n}}^{m,n}\}$ be a non-homogeneous Poisson process of rate $mp_0+np_1$. Following the same arguments in \cite{HP}, assign a mark from the set $\{1,2\}$ to each point of $\mathcal{Z}'_{m,n}$. Specifically, a point $\bx$ is assigned mark $1$ with probability $mp_0(\bx)\big/\big(mp_0(\bx)+np_1(\bx)\big)$ and mark $2$ otherwise. Let $\widetilde{\mathcal{X}}_m$ and $\widetilde{\mathcal{U}}_n$ be the set of points of $\mathcal{Z}'_{m,n}$ marked $1$ and $2$ respectively. Also denote $\widetilde{\mathcal{A}}_{m,n}$ the cross match statistic given from optimal weighted matching over $\widetilde{\mathcal{X}}_m\cup \widetilde{\mathcal{U}}_n$. Define the probability of two points in $\mathcal{Z}'_{m,n}$ having different marks by $g_{m,n}(\bx,\by)$: 
\begin{equation}
g_{m,n}(\bx,\by)=\diy\frac{mp_0(\bx)np_1(\by)+np_1(\bx) m p_0(\by)}{(mp_0(\bx)+np_1(\bx))(mp_0(\by)+np_1(\by))}.
\end{equation}
We know that $m/N\rightarrow c_0$ and $n/N\rightarrow c_1$, hence
$g_{m,n}(\bx,\by)\rightarrow g(\bx,\by)$ where 
\begin{equation}
g(\bx,\by)=\diy \frac{c_0 c_1\big(p_0(\bx)p_1(\by)+p_1(\bx)p_0(\by)\big)}{\big(c_0p_0(\bx)+c_1p_1(\bx)\big)\big(c_0p_1(\by)+c_1p_1(\by)\big)}.
\end{equation}
So, the conditional expectation $E\big[\widetilde{\mathcal{A}}_{m,n}|\mathcal{Z}'_{m,n}]$ becomes:
\begin{equation}\label{thm1:eq11}
\mathop{\sum\sum}\limits_{1\leq i<j\leq W_{m,n}} g_{m,n}(\BZ_i^{m,n},\BZ_j^{m,n})\mathbf{1}\big\{(\BZ_i^{m,n},\BZ_j^{m,n})\in M^*(\mathcal{Z}'_{m,n})\big\}.
\end{equation}
By taking expectations in (\ref{thm1:eq11}), one yields $
E\big[\widetilde{\mathcal{A}}_{m,n}\big]$.\\
Let $\mathcal{Z}_{m,n}:=\big\{\BZ_1^{m,n},\BZ_2^{m,n},\dots, \BZ_{m,n}^{(m+n)}\big\}$ be the original non-Poissonized set of points. 
By the fact that 
$$E\big[|M_m+N_n-(m+n)|\big]=o(m+n),$$
the Poissonized limit of $E\big[\tilde{\mathcal{A}}_{m,n}\big]$.  
Set $p(\bx)=c_0p_0(\bx)+c_1p_1(\bx)$, then $p_{m,n}(\bx)\rightarrow p(\bx)$. Using Lemma \ref{lem2}, we get
\begin{equation}\begin{array}{l}
\diy\frac{E\big[\widetilde{\mathcal{A}}_{m,n}\big]}{(m+n)}\rightarrow \diy \frac{1}{2}\int_{\mathcal{R}^d}g(\bx,\bx) p(\bx)\\
\\
\qquad=\diy c_0\;c_1\;\diy\int_{\mathcal{R}^d}\frac{p_0(\bx)p_1(\bx)}{c_0p_0(\bx)+c_1p_1(\bx)}.
\end{array}\end{equation}
This completes the proof of Theorem \ref{thm1}. \hfill$\square$

\section{Experiments}
\label{sec:experiments}
We perform multiple experiments to demonstrate the utility of the proposed direct estimator of HP-divergence in terms of dimension and sample size. We subsequently apply our estimator to determine empirical bounds on the Bayes error rate for various datasets. 

For the following simulations, the sample sizes for each class were equal ($m=n$). Each simulation used a multivariate Normal distribution for each class.

We first analyze the estimator's performance as the sample size $N = m + n$ increases. For each value of $N$, the simulation was run 50 times, and the results were averaged. Samples from each class were i.i.d. 2-dimensional Normal random variables, with $\mu_0 = [0,0]$ and $\mu_1 = [1,1]$, $\Sigma_0 = \Sigma_1 = I_2$. 

\begin{figure}[h]
  \includegraphics[width=\linewidth]{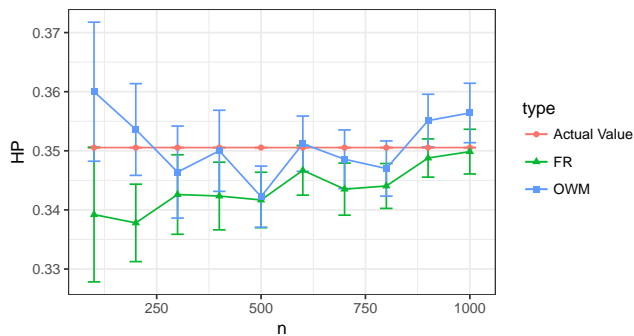}
  \caption{HP-divergence estimation vs. sample size $n$. Error bars denote the standard deviation over 50 trials. The proposed estimator and the FR estimator perform approximately equivalently over this range of sample sizes.}
  \label{fig3}
 \end{figure}
 We see that as $N$ increases the performance of the FR estimator and our proposed estimator (labeled OWM) are comparable for $N$ up to 1000. The observed variance of our estimators are slightly higher than the FR estimator. For dimension $d=2$ this is not surprising as we would expect the FR estimator to perform the best in this case. 

\begin{figure}[h]
  \includegraphics[width=\linewidth]{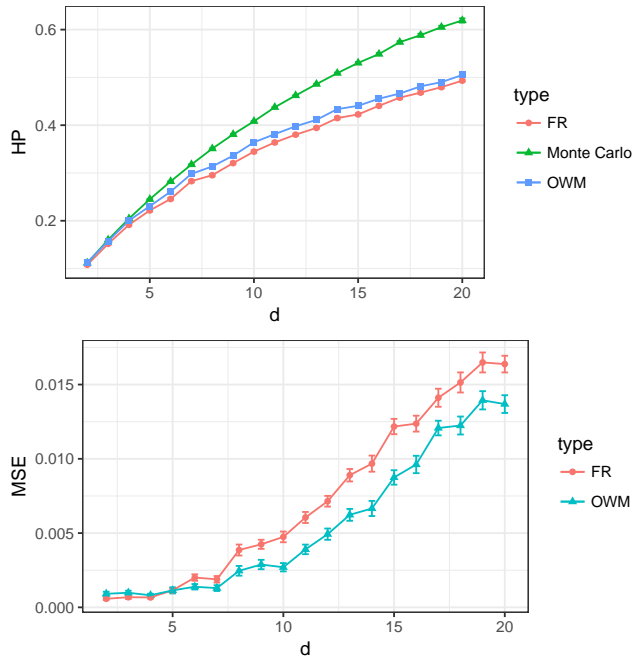}
  \caption{HP-divergence (top) and empirical MSE (bottom) vs. dimension. The empirical MSE of both estimators increases for larger dimensional data sets. The MSE is better for the proposed (OWM) estimator.}
  \label{fig4}
 \end{figure}

Figure~\ref{fig4} (top) shows the averaged estimates of the HP-divergences over increasing dimension. Here we see that the proposed cross-matching estimator shows improvement with respect to the FR estimator, as expected. For each dimension evaluated in Figure~\ref{fig4}, $N = 1000$, and $\mu_0 = [0]_d$ and $\mu_1 = [0.5]_d$, $\Sigma_0 = \Sigma_1 = I_d$. The proposed cross-matching estimator is slightly less biased as dimension increases, and as shown in Figure~\ref{fig4} (bottom) we improve in empirical MSE.

Next we show the results of applying the HP-divergence estimator to 4 different real data sets. Table 1 shows the cross match statistics and estimated upper bounds for Bayes Error (denoted by the column labeled $\epsilon$).

\begin {table}[H]
\begin{center}
\begin{tabular}{ |c||c|c|c|c|c|  }
\hline
 \multicolumn{6}{|c|}{Bayes Error Bounds} \\
\hline
Data set & $\mathcal{A}(\mathcal{X}_N)$  & $\widehat{D}_c$ & $n$ & $m$& $\epsilon$ \\
 \hline
Breast cancer~\cite{Wolberg1990} & 33  & 0.791   & 488 &  241&  0.093\\
Mines vs. Rocks~\cite{Lichman2013} & 7  &   0.864  &  97   & 111&  0.067\\
Pima diabetes~\cite{Lichman2013} & 67 & 0.641 & 549 &  283 & 0.161\\
Hyper thyroid~\cite{Lichman2013} & 37  & 0.743 & 3012 &  151&   0.023\\
 \hline
\end{tabular}
\caption {{$\mathcal{A}(\mathcal{X}_N)$, $\widehat{D}_c$, $n$, $m$ and $\epsilon$ are the cross-match statistics, HP-divergence estimates using $\mathcal{A}(\mathcal{X}_N)$, sample sizes and upper bounds for Bayes Error respectively.} }
\end{center}
\label{tab:real_data}
\end{table}
\section{Conclusion}
\label{sec.conclusion}

We proposed a new dimension-independent direct estimator of HP-divergence using a statistic derived from optimal weighted matching. The estimator is more accurate than the FR approach and its variance is independent of the dimension of the support of the distributions. This translates to improved MSE performance as compared to other HP-divergence estimation methods, especially for high dimension. We validated our proposed estimator using simulations, and illustrated the approach for the meta-learning problem of estimating Bayes classification error for four real-world data sets.

\section{Acknowledgements}
We thank the UCI machine learning repository for the use of their various datasets used in this paper~\cite{Lichman2013}.

\label{sec:refs}
\bibliographystyle{IEEEbib}
\bibliography{refs}
\newpage


\end{document}